\documentclass[prd,aps,10pt,notitlepage,floatfix,twocolumn]{revtex4-1}
\pdfoutput=1
\usepackage{amssymb}
\usepackage{amsmath}    
\usepackage{graphicx}   
\usepackage{verbatim}   
\usepackage{color}      
\usepackage{hyperref}   
\usepackage{array}
\newcolumntype{C}[1]{>{\centering\arraybackslash}m{#1}}
\usepackage{float}
\begin{document}

\title{Constraints on the Nambu-Goto cosmic string contribution to the CMB power spectrum in light of new temperature and polarisation data}
\author{Andrei Lazanu}
\email{A.Lazanu@damtp.cam.ac.uk}
\affiliation{Centre for Theoretical Cosmology, DAMTP, University of Cambridge, CB3 0WA, UK}
\author{Paul Shellard}
\email{E.P.S.Shellard@damtp.cam.ac.uk}
\affiliation{Centre for Theoretical Cosmology, DAMTP, University of Cambridge, CB3 0WA, UK}
\date{\today}

\begin{abstract}
Cosmic strings generate vector and tensor modes in the B-channel of polarisation, as well as the usual temperature power spectrum and E-mode polarisation spectrum. We use the power spectrum obtained from high-resolution Nambu-Goto cosmic string simulations together with the Planck and BICEP2 likelihoods to explore the degeneracies appearing between cosmic strings and other cosmological parameters in different inflationary scenarios, as well as the constraints that can be imposed on cosmic strings in each of these situations. In standard $\Lambda$CDM, the Planck likelihood yields an upper limit $G\mu<1.49 \times 10^{-7}$ (95\% confidence). We also analyse the possibility of explaining the BB power spectrum signal recently detected by the BICEP2 probe. We find that cosmic strings alone are able to explain only part of the B-mode polarisation signal. Apart from the standard $\Lambda$CDM model, we look at the following non-minimal parameters: the running of the spectral index, non-zero tensor-to-scalar ratio, additional degrees of freedom ($N_{eff}$) and sterile neutrinos. We find that in both Planck and BICEP2 scenarios adding $N_{eff}$ induces degeneracies between cosmic strings and $N_{eff}$ and other $\Lambda$CDM parameters. With $N_{eff}$ a larger contribution from cosmic strings is allowed, even favoured, but after combining with large-scale structure data, such as BAOs, strings remain strongly constrained.
\end{abstract}

\maketitle

\section{Introduction}
Recent B-mode polarisation results \cite{bicep} potentially open a new window on the Universe, especially if the signal has a primordial origin, though this is yet to be confirmed \cite{2014arXiv1409.5738P}. Such a signal is not explained by the standard $\Lambda$CDM and would require extensions by adding extra parameters.

One of the simplest additions to the minimal $\Lambda$CDM is primordial tensor modes generated by inflation at a high energy scale. The BICEP2 Collaboration estimates a tensor-to-scalar ratio around  $r=0.20$ ($r=0.16$ after foreground subtraction), but the results are in tension with the standard $\Lambda$CDM model which is also used in Planck papers \cite{planckres}. This could suggest that additional degrees of freedom are required in order to relieve this tension \cite{bicep}, possibly by allowing for a scale-dependent spectral index, the running of the spectral index ($n_{run}=dn_s/dlnk$). Many different cosmological scenarios have been proposed to relieve this tension: the curvature of the universe \cite{freivogel}, the number of effective degrees of freedom, the sum of the neutrino masses, the Helium abundance and a sterile neutrino mass \cite{rosner}, running of the spectral index and dust \cite{PhysRevD.90.103504}.

A different type of solution to explain the B-mode polarisation is a signal from topological defects. Defects generate both vector and tensor modes in the B channel \cite{cmbact, wyman, PhysRevD.76.043005} and hence they are a natural candidate for reconciling the tension between the datasets. Many inflationary models involve a phase transition in the early universe where cosmic strings can be formed naturally \cite{shbook, 2014arXiv1403.6688S}. Both the groups working on strings generated through the phenomenological unconnected segment model (USM) \cite{cmbact, albrecht, Pogosian, copeland} and the Abelian-Higgs cosmic strings \cite{hindmarsh2006, kunz-cosmomc, PhysRevD.90.103504} have analysed the possibility of using cosmic strings to explain the polarisation signal form the BICEP2 probe \cite{2014arXiv1403.4924L, 2014arXiv1403.6105M}. These authors have evaluated the implications of the presence of cosmic strings and have concluded that they cannot alone explain the whole signal in the B-mode polarisation, though they could make some contribution. We re-examine these conclusions using the improved estimates of the string CMB power spectrum calculated directly from Nambu-Goto simulations \cite{string-paper1}.

After a short introduction, the paper is split into two parts. In the first part of this paper we analyse the contribution of Nambu-Goto cosmic strings obtained from simulations \cite{Allen, landriau2003} to the power spectrum in different scenarios prior to the release of the BICEP2 data, using Planck and WMAP polarisation data. We look at the degeneracies between cosmic strings and other cosmological parameters and we also observe the influence of SPT/ACT \cite{0004-637X-743-1-90, Dunkley:2013vu} and Baryon Acoustic Oscillations (BAO) in modified the allowed contribution from cosmic strings as well as to reduce degeneracies. We introduce non-minimal parameters such as $r$, the running of the spectral index (running),  the number of effective degrees of freedom $N_{eff}$ and also the mass of some sterile neutrinos

In the second part of the paper we analyse the contribution of cosmic strings in light of a significant tensor contribution measured in the BICEP2 data. We analyse degeneracies in similar situations to the Planck counterpart and we also analyse the possibility of explaining the B-mode signal by the presence of cosmic strings on their own and also using additional non-minimal parameters as in the previous section.

\section{Cosmic string power spectrum}

In order to evaluate the cosmic string power spectrum, we have used three high-resolution Nambu-Goto cosmic string simulations based on the Allen and Shellard code \cite{Allen}. The simulations cover the entire cosmological time from the end of inflation, through the radiation and matter eras, until late into the cosmological constant epoch. Each of the simulations consists of the discrete time evolution of a network of strings, starting with Vachaspati-Vilenkin initial conditions \cite{vilenkin, shbook}. At every step, the network is interpolated on a three-dimensional grid in real space and the energy-momentum tensor of the system is evaluated at each grid vertex. The stress-energy tensor is then transformed into Fourier space and a scalar-vector-tensor decomposition is performed.

In order to evaluate the cosmic string power spectrum, one starts by considering the first order perturbations to the Einstein Equation:
\begin{equation}
\delta G_{\mu\nu}+\Lambda \delta g_{\mu\nu}=8\pi G \delta T_{\mu\nu}
\label{einstein}
\end{equation}
where the perturbation to $T_{\mu\nu}$ is due to the usual matter perturbations and the cosmic string energy-momentum tensor. The LHS of the equation is not modified by cosmic strings and the metric $g_{\mu\nu}$ is a small perturbation around the FLRW metric. By splitting this equation into scalar, vector and tensor parts, one obtains the perturbation equations for the metric sourced by cosmic strings in Fourier space. As the cosmic string perturbations are uncorrelated with the primordial fluctuations, the cosmic string power spectrum can be separately evaluated by choosing correct initial conditions. The scalar, vector and tensor equations decouple and hence the Boltzmann equations can be treated separately for each of them. These equations (without cosmic strings) have been implemented into the Boltzmann solver CMBFAST \cite{seljak} and we have modified this code to accommodate the cosmic string sources.

The Boltzmann equations are scalar equations and the string energy momentum tensor depends on the magnitude as well the direction of the wave-vector. The complexity of the problem and the amount of information make it impossible to integrate directly over the three spatial directions. This is however not necessary. As cosmic strings are active sources, it has been shown \cite{durrer} that to first order in perturbation theory the whole information is contained into the unequal time correlators (UETCs), which are the two-point correlation functions of different components of the energy momentum tensor. Using the set of equations implemented in CMBFAST, one only needs the following five unequal time correlators: $\langle \Theta_{00}\Theta_{00} \rangle$, $\langle \Theta^S\Theta^S \rangle$, $\langle \Theta_{00}\Theta^S \rangle$, $\langle \Theta^V\Theta^V \rangle$ and $\langle \Theta^T\Theta^T \rangle$, all the other sets of correlators made out of these stress-energy tensor components being 0. Assuming the string network is scaling in time, UETCs appear only as functions of $k\tau$:
\begin{equation}
\sqrt{\tau\tau'}\langle\Theta_{a}(\textbf{k},\tau)\Theta_{b}(-\textbf{k},\tau')\rangle=C_{ab}(k\tau,k\tau')
\end{equation}

They are positive definite matrices, and hence can be diagonalised:
\begin{equation}
C_{ab}(k\tau,k\tau')=\sum_i\lambda_iv_a^{(i)}(k\tau)v_b^{(i)T}(k\tau')
\end{equation}

\begin{figure}[!h]	
\begin{center}$
\begin{array}{c}
\includegraphics[width=2.8in]{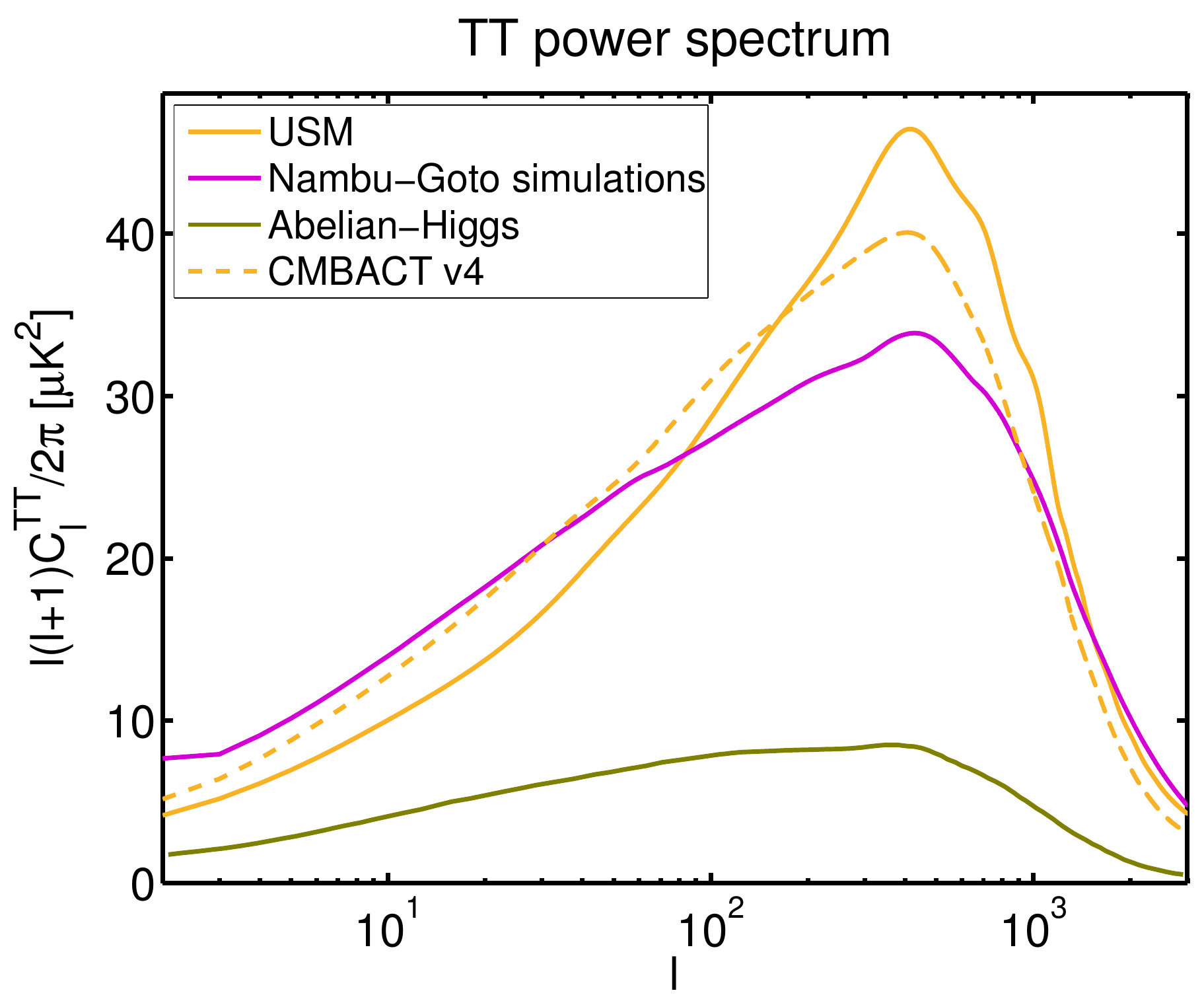} \\ 
\includegraphics[width=2.8in]{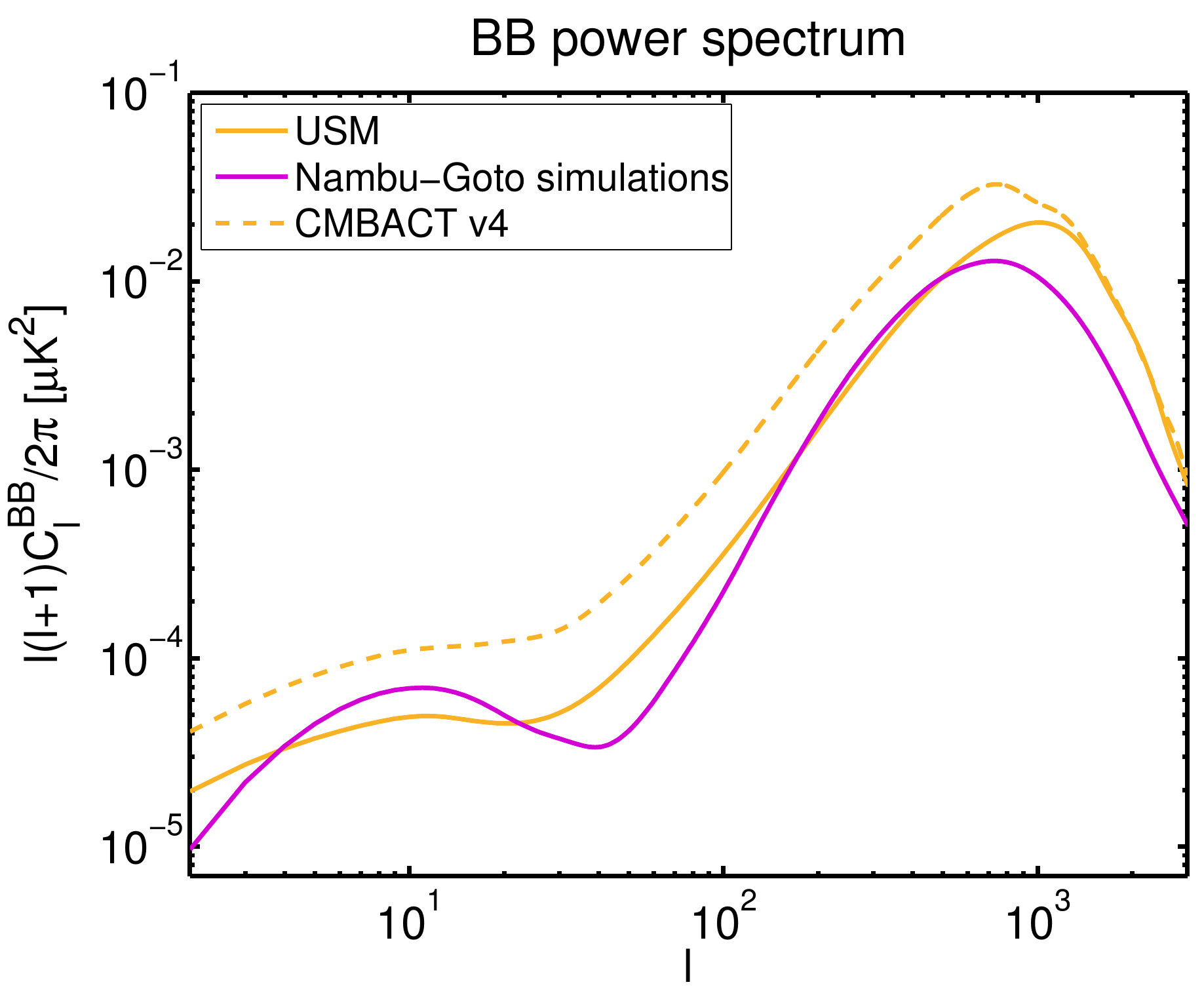}
\end{array}$
\caption{TT (top) and BB (bottom) cosmic string power spectra corresponding to $G\mu/c^2=1.5 \times 10^{-7}$}
\label{TTBB}
\end{center}
\end{figure}

The eigenmodes are coherent, and hence the total angular power spectrum can be obtained as a sum of the contributions from each eigenvector from the scalar, vector and tensor parts separately
\begin{equation}
C_l^{\text{string}}=\sum_{S,V,T}\sum_i \lambda_i C_l^{(i)}
\label{sumcls}
\end{equation}

\begin{figure*}[!htb]	
\begin{center}$
\begin{array}{c}
\includegraphics[width=6in]{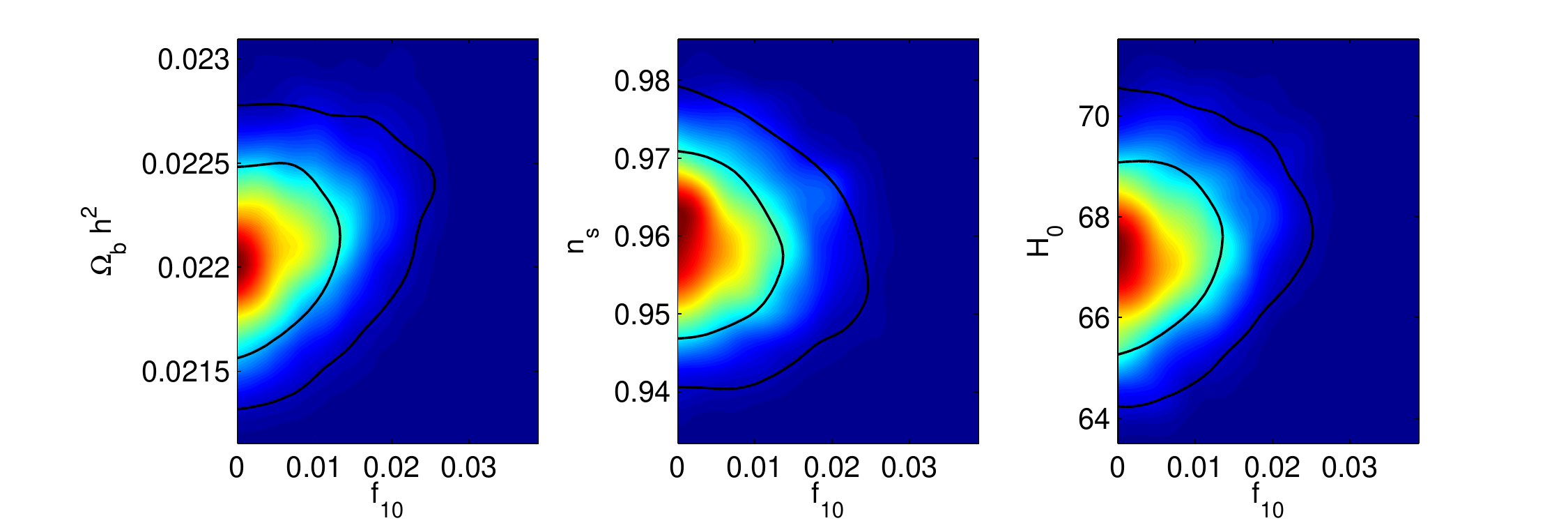}
\end{array}$
\caption{Marginalised likelihoods in the Planck + WP \& strings model}
\label{planck_direct}
\end{center}
\end{figure*}

Hence, in order to implement this method, we just have to substitute the energy-momentum tensor component from cosmic strings with its corresponding eigenvector in the Boltzmann equations:
\begin{equation}
\Theta(\textbf{k},\tau) \to \frac{v^{(i)}(k\tau)}{\sqrt{\tau}}
\label{theta2v}
\end{equation}

The procedure summarised above assumes scaling throughout the history of the universe. This is however not true, as we don't obtain the same result using the 3 simulations separately. Hence, we devised a method to incorporate them at the same time. After calculating the eigenvectors from the UETCs of each simulations, we assume its validity only during the time interval used to generate it and we make the eigenvectors decay outside it. Then we calculate the power spectrum created in this way.For example, in the radiation era:
\begin{equation}
\Theta(\textbf{k},\tau) \to \left\{
\begin{array}{cc}
\frac{v_{\text{radiation}}(k\tau)}{\sqrt{\tau}} & \text{if } \tau \in \text{radiation era} \\
0 & \text{if } \tau \not\in \text{radiation era}\\
\end{array} \right.
\label{impartire}
\end{equation}

The final $C_l$'s are obtained as sums of $C_l$'s from the individual simulations:
\begin{equation}
C_l^{\text{string}}=\sum_j\sum_{S,V,T}\sum_i \lambda_i C_l^{(i)s_j}
\end{equation}
where $s_1$, $s_2$, $s_3$ are the 3 simulations.

The details of this procedure are explained in \cite{string-paper1}.

In Figure \ref{TTBB} we show the TT and BB power spectra that we have obtained from these simulations, together with the unconnected segment model  \cite{Pogosian, albrecht} results and the spectra obtained from a revised version of CMBACT (dashed line) \cite{cmbact, pogosian-arizona}. For the temperature spectrum we also show the Abelian-Higgs result form the Planck Collaboration paper \cite{planckstr}.
The USM plot corresponds to CMBACT version 3, with default values of the parameters ($v= 0.65$, $\alpha= 1.9$, $\xi=0.13$), as used in the Planck paper. The CMBACT version 4 uses a different model for the USM,  ($v=0.65$, $\alpha=1$, $\xi=0.15$), where these parameters are only used as initial conditions in the Boltzmann equations and hence the results depend only weakly on them. The difference between the two CMBACT results is due to this change of model, and also to a bug in the older version of the code which overestimated the contribution of vector modes by a factor of two.

\section{Planck case}

We have added the power spectrum of the cosmic strings to the inflationary one and we have used the full Planck likelihoods and WMAP polarisation data \cite{planckres} to analyse the contribution to the CMB power spectrum from Nambu-Goto cosmic strings with Markov chain Monte Carlo methods with the COSMOMC code \cite{cosmomc, cosmomc2}. This method involves evaluating the power spectrum each time the parameters are modified, by calling an instance of the code CAMB \cite{CAMB}. The total power spectrum is obtained from the sum between the inflationary spectrum and the cosmic strings spectrum, because the cosmic string sources, which are active sources, are uncorrelated with the primordial perturbations \cite{wyman}. This would in principle require the calculation of the cosmic string power spectrum many thousands of times, for each choice of cosmological parameters, which is not feasible because calculating the cosmic string power spectrum by itself requires several CPU hours of computational work. It has been shown \cite{Pogosian, albrecht} that it evolves much slower as a function of the parameters compared to its inflationary counterpart. Also, the cosmic strings are expected to contribute less than 5\% in the total power spectrum, so as the cosmological parameters are varied in the allowed regions, the string power spectrum does not vary more than 20\% \cite{kunz-cosmomc, battye-cosmomc}. This gives overall better than 1\% accuracy for the contribution of cosmic strings, which is greater than the accuracy of CAMB.

\begin{figure*}[!htb]	
\begin{center}$
\begin{array}{c}
\includegraphics[width=6in]{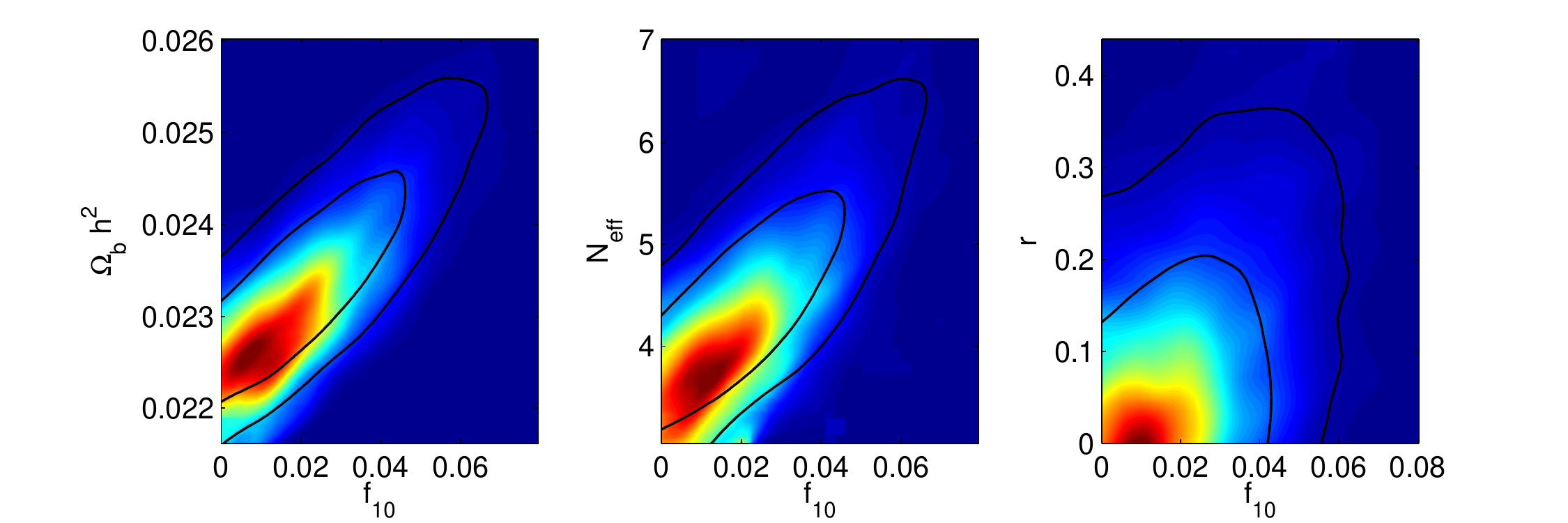}
\end{array}$
\caption{Marginalised likelihoods obtained when adding cosmic strings, $N_{eff}$ and tensor modes ($r$)}
\label{neff_r_planck2}
\end{center}
\end{figure*}

Cosmic strings are quantified through the parameter $f_{10}$ which represents the fractional power spectrum due to cosmic strings at the 10th multipole \cite{battye2-cosmomc, planckstr} and is related to the cosmic string tension $G\mu/c^2$. We have added this to the base model, called $\Lambda$CDM, which is based on the following six parameters: the baryon density $\Omega_b h^2$, the cold dark matter density $\Omega_c h^2$, the Thomson optical depth to recombination $\tau$, the proxy for the angular acoustic scale at recombination $\theta_{MC}$, the amplitude of the initial curvature power spectrum (at \textit{k}=0.05 Mpc$^{-1}$) and the scalar spectral index $n_s$.

In this simplest case, we have found the constraints: $G\mu<1.49 \times 10^{-7}$ and $f_{10}<0.0193$ at 95\% confidence level, which are comparable with the results obtained by the Planck Collaboration \cite{planckstr}. The improvement in the fit after including cosmic strings is small. There are few degeneracies with cosmic strings, and $\Lambda$CDM parameters change very little after the introduction of cosmic strings. In Figure \ref{planck_direct} we plot the marginalised likelihoods in the $f_{10}-\Omega_bh^2$, $f_{10}-H_0$ and $f_{10}-n_s$ planes. Here, the two-dimensional plot is similar to Figure 10 of Ref. \cite{planckstr}. The constraint that we have obtained is slightly stronger compared to the Planck one ($G\mu/c^2<1.5 \times 10^{-7}$) \cite{planckstr}. We have validated our formalism by obtaining the constraint for the unconnected segment model power spectrum. Therefore, we have concluded that this is due to the fact that our power spectrum has a different shape and more power at low multipoles.

Apart from this scenario we have also considered adding different non-minimal parameters to the model and we have looked at the degeneracies that appeared. The parameters we have considered adding are the following:
\begin{itemize}
 \item $r$, which is the tensor-to-scalar ratio evaluated at $k=0.002$ Mpc$^{-1}$. This also enforces the relation $n_t=-r/8$ for the tensor spectral index $n_t$
 \item \textit{running} of the spectral index
 \item $N_{eff}$, the effective number of neutrino-like relativistic degrees of freedom. The minimal case corresponds to $N_{eff}$=3.046 and the additional degrees of freedom are quantised by the parameter $\Delta N_{eff}=N_{eff}-3.046$
 \item $m_{\nu,{\rm{sterile}}}^{\rm{eff}}$, the mass of a \textit{sterile neutrino}. The sterile neutrinos are motivated by the discovery of neutrino oscillations (e.g. Ref. \cite{neutrino-mass-lesgourgues})
\end{itemize}

\begin{figure}[!htb]	
\begin{center}$
\begin{array}{c}
\includegraphics[width=3in]{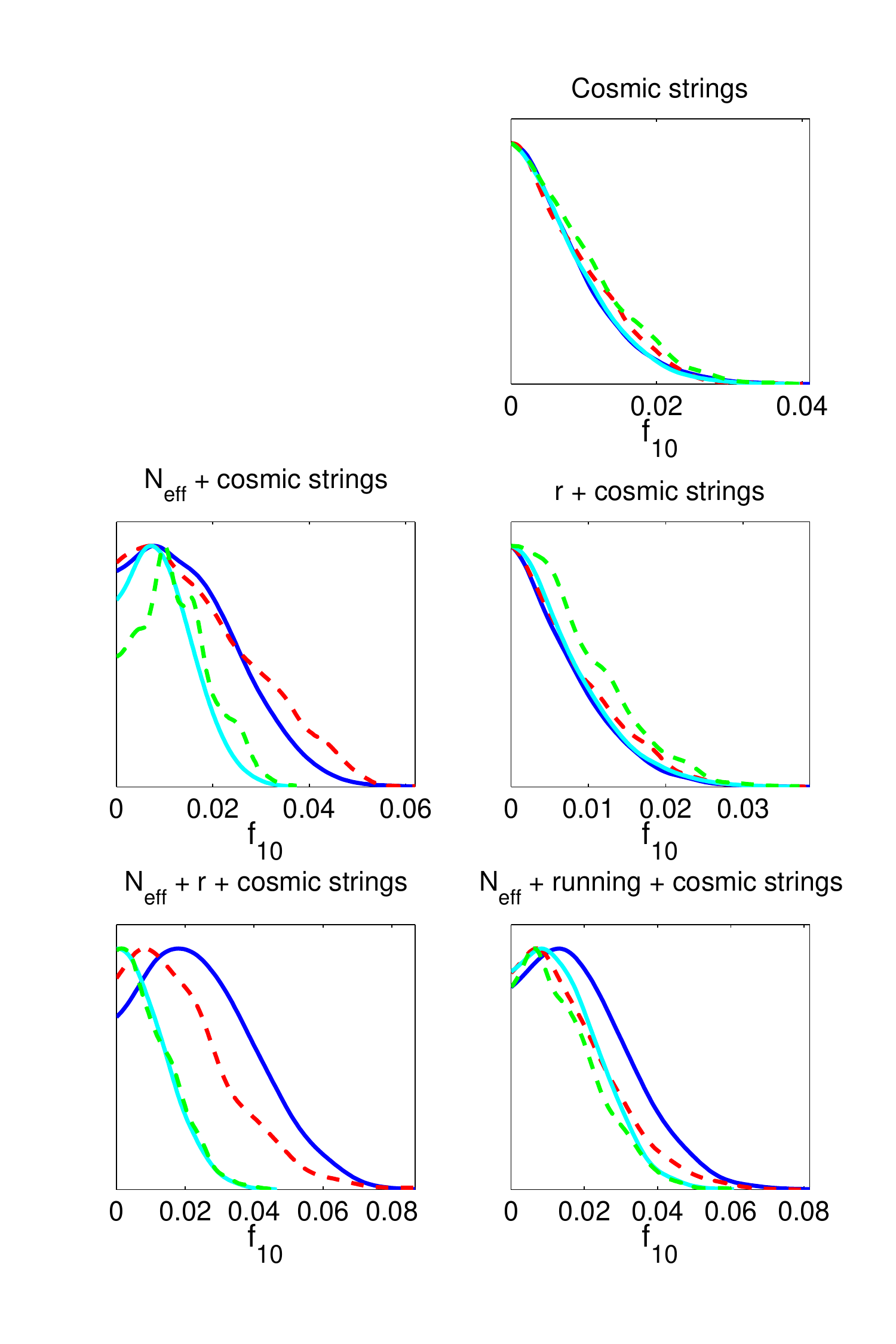} 
\end{array}$
\caption{Mean likelihoods of the samples (red dotted lines) and marginalised probabilities (blue solid lines) for parameter $f_{10}$ in the following situations (from left to right and top to bottom): Planck \& strings; Planck \& strings \& $N_{eff}$; Planck \& strings \& $r$; Planck \& strings \& $N_{eff}$ \& $r$, Planck \& $N_{eff}$ \& running \& cosmic strings. The green and cyan curves respectively represent the mean likelihoods and marginalised probabilities of the samples after the introduction of HighL \& BAO.}
\label{planck_1dlike}
\end{center}
\end{figure}

The most interesting results that we have are shown in Table \ref{params_planck}. The other scenarios are tabled in the Appendix (Table \ref{params_planck_appendix}).

\begin{table*}[p!]
\centering
\caption{Values of the cosmological parameters when considering only the Planck likelihoods, with $1\sigma$ error levels full likelihood analysis (all cases also include the Planck nuisance parameters)}
\begin{tabular}{|c|C{3cm}|C{3cm}|C{3cm}|C{3cm}|}
\hline
Parameter                           & $\Lambda$CDM        & strings            & strings, $n_{run}$     & strings, $r$, $N_{eff}$       \\ \hline
$G\mu/c^2 < (2\sigma)$              & -                   & $1.49 \times 10^{-7}$  & $1.88 \times 10^{-7}$   & $2.49 \times 10^{-7}$ \\ \hline
$G\mu/c^2$ (best fit)               & -                   & $4.99 \times 10^{-8}$  & $8.23 \times 10^{-8}$   & $1.09 \times 10^{-8}$ \\ \hline
$n_{run}$                           & -                   & -                  & $-0.020 \pm 0.010$& -                   \\ \hline
$r$                                 & -                   & -                  & -                   & $0.12 \pm 0.09$                    \\ \hline
$\Delta N_{eff}$                    & -                   & -                  & -                   & $1.574 \pm 0.748$      \\ \hline 
$H_0$                               & $67.20\pm 1.16$     & $67.42 \pm 1.20$   & $67.46 \pm 1.22$    & $80.59 \pm 6.57$     \\ \hline 
$100\Omega_b h^2$                   & $2.202\pm 0.027$    & $2.209 \pm 0.029$  & $2.237 \pm 0.034$   & $2.354 \pm 0.077$    \\ \hline
$\Omega_c h^2$                      & $0.120 \pm 0.003$    & $0.119 \pm 0.003$ & $0.120 \pm 0.003$   & $0.135 \pm 0.008$     \\ \hline
$\tau$                              & $0.089 \pm 0.013$   & $0.087 \pm 0.013$  & $0.098 \pm 0.016$   & $0.100 \pm 0.016$   \\ \hline
$100\theta_{MC}$                    & $1.0412 \pm 0.0006$ & $1.0412 \pm 0.0006$& $1.0413 \pm 0.0007$ & $1.0402 \pm 0.0007$    \\ \hline
$ln(10^{10}A_s)$                    & $3.088 \pm 0.025$   & $3.078 \pm 0.026$  & $3.101 \pm 0.032$   & $3.123 \pm 0.033$  \\ \hline
$n_s$                               & $0.959 \pm 0.007$   & $0.958\pm 0.007$   & $0.952 \pm 0.008$   & $1.017 \pm 0.027$    \\ \hline
$-ln\mathcal{L}$                    & 4902.4              & 4902.1             & 4902.0              & 4902.5     \\ \hline \hline

Parameter                           & $N_{eff}$ (Planck) & strings, $N_{eff}$ (Planck) & strings, $N_{eff}$ (Planck + HighL)& strings, $N_{eff}$ (Planck + HighL + BAO)  \\ \hline
$G\mu/c^2 < (2\sigma)$              & -                  & $2.28 \times 10^{-7}$& $1.80 \times 10^{-7}$& $1.58 \times 10^{-7}$ \\ \hline
$G\mu/c^2$ (best fit)               & -                  & $7.35 \times 10^{-8}$& $1.77 \times 10^{-7}$& $1.34 \times 10^{-7}$\\ \hline
$\Delta N_{eff}$                    & $0.563 \pm 0.316$  & $1.072 \pm 0.564$    & $1.186 \pm 0.528 $   & $0.658 \pm 0.304$\\ \hline 
$H_0$                               & $71.34 \pm 2.66$   & $75.96 \pm 4.84$     & $76.46 \pm 4.45$     & $71.51 \pm 1.95$\\ \hline 
$100\Omega_b h^2$                   & $2.243 \pm 0.037$  & $2.305 \pm 0.062$    & $2.302 \pm 0.054 $   & $2.243 \pm 0.030$\\ \hline
$\Omega_c h^2$                      & $0.127 \pm 0.0047$ & $0.132 \pm 0.006$    & $0.134 \pm 0.07$     & $0.129 \pm 0.05$\\ \hline
$\tau$                              & $0.095 \pm 0.015$  & $0.098 \pm 0.015$    & $0.097 \pm 0.015$    & $0.090 \pm $0.013\\ \hline
$100\theta_{MC}$                    & $1.0405 \pm 0.0007$& $1.0404 \pm 0.0007$  & $1.0402 \pm 0.0007$  & $1.0404 \pm 0.0007$\\ \hline
$ln(10^{10}A_s)$                    & $3.117 \pm 0.031$  & $3.117 \pm 0.034$    & $3.121 \pm 0.033$    & $3.104 \pm 0.029$\\ \hline
$n_s$                               & $0.980 \pm 0.014$  & $0.996 \pm 0.020$    & $0.997 \pm 0.019$    & $0.976 \pm 0.010$          \\ \hline 
$-ln\mathcal{L}$                    &  4902.0            & 4902.6               & 5255.3               & 5256.1    \\ \hline  \hline

Parameter                           & $n_{run}$ (Planck)     & $n_{run}$, $N_{eff}$, strings (Planck) & $n_{run}$, $N_{eff}$, strings (Planck + HighL) &  $n_{run}$, $N_{eff}$, strings (Planck + HighL + BAO)    \\ \hline
$G\mu/c^2 < (2\sigma)$              & -                   & $2.28 \times 10^{-7}$& $2.06 \times 10^{-7}$& $1.95 \times 10^{-7}$ \\ \hline
$G\mu/c^2$ (best fit)               & -                   & $1.03 \times 10^{-7}$& $1.75 \times 10^{-7}$& $3.57 \times 10^{-8}$ \\ \hline
$n_{run}$                           & $-0.015 \pm 0.009$  & $-0.054\pm 0.015$  & $-0.008\pm 0.015$   &$-0.014\pm 0.011$        \\ \hline
$\Delta N_{eff}$                    & -                   & $0.935 \pm 0.713$  & $0.969 \pm 0.733$   & $0.386 \pm 0.289$      \\ \hline 
$H_0$                               & $67.00 \pm 1.20$    & $74.84 \pm 5.87$   & $74.73 \pm 5.92$    & $70.21 \pm 1.87$    \\ \hline 
$100\Omega_b h^2$                   & $2.215 \pm 0.030$   & $2.300 \pm 0.062$  & $2.294 \pm 0.058$   & $2.258 \pm 0.032$    \\ \hline
$\Omega_c h^2$                      & $0.121 \pm 0.027$   & $0.130 \pm 0.008 $ & $0.131 \pm 0.008$   & $0.125 \pm 0.005$     \\ \hline
$\tau$                              & $0.097 \pm 0.015$   & $0.099 \pm 0.016$  & $0.099\pm 0.016$    & $0.097 \pm 0.015$    \\ \hline
$100\theta_{MC}$                    & $1.0412 \pm 0.0006$ & $1.0405 \pm 0.0008$& $1.0404\pm 0.0008$  & $1.0408 \pm 0.0007$   \\ \hline
$ln(10^{10}A_s)$                    & $3.108 \pm 0.030$   & $3.117 \pm 0.034$  & $3.122\pm 0.033$    & $3.108 \pm 0.031$   \\ \hline
$n_s$                               & $0.954 \pm 0.008$   & $0.989 \pm 0.028$  & $0.986 \pm 0.028$   & $0.967 \pm 0.012$               \\ \hline
$-ln\mathcal{L}$                    & 4901.7              & 4902.2             & 5255.6              & 5257.8              \\ \hline  

\end{tabular}
\label{params_planck}
\end{table*}

\begin{figure*}[!p]	
\begin{center}$
\begin{array}{cc}
\includegraphics[width=0.77in]{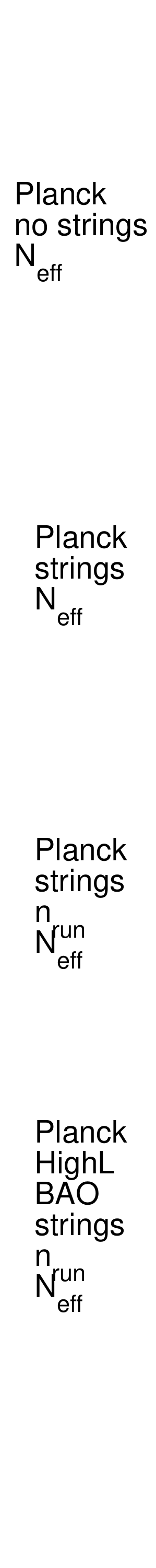}
\includegraphics[width=6.05in]{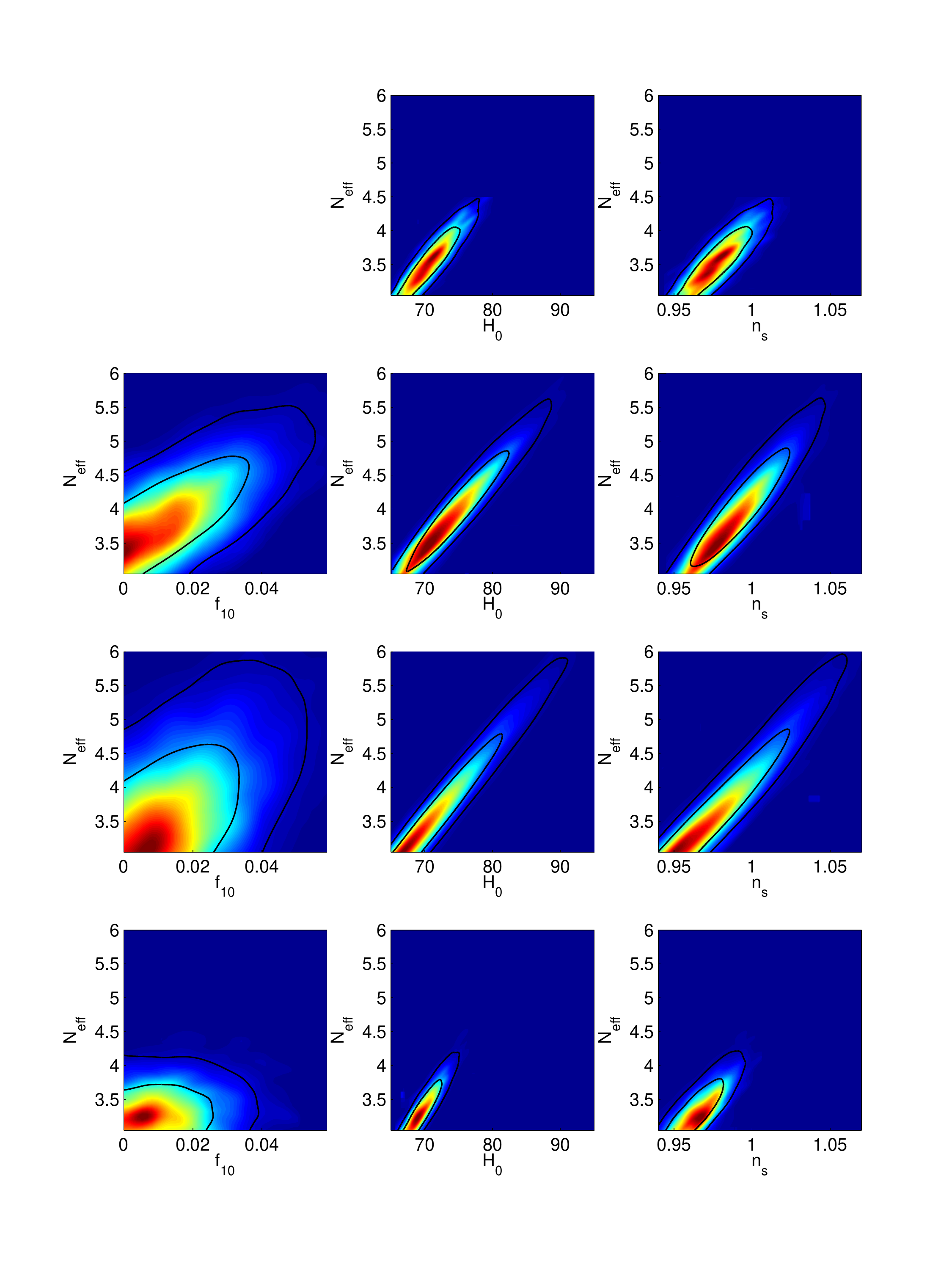}
\end{array}$
\caption{Two-dimensional marginalised likelihoods in the $f_{10}$-$N_{eff}$, $H_0$-$N_{eff}$ and $n_s$-$N_{eff}$ planes in the following cases (top to bottom): $N_{eff}$ only (no strings), $N_{eff}$ and cosmic strings, $N_{eff}$, running and cosmic strings, $N_{eff}$, running and cosmic strings, with SPT/ACT and BAOs}
\label{planck_2d}
\end{center}
\end{figure*}

We have observed that $N_{eff}$ is in most cases very degenerate with cosmic strings and hence it allows it to attain huge values. For example, looking just at the second row of Table \ref{params_planck}, we observe that the preferred values for $N_{eff}$ increase after adding cosmic strings, to $\Delta N_{eff}>1$. The error bar also increases, suggesting the fact that there is a degeneracy appearing after adding the parameter $f_{10}$. The error bar also increases considerably on the baryon contribution, $H_0$ and $n_s$. The value of the Hubble constant is increased massively from the $\Lambda$CDM result. We have explored ways in order to fix this, by adding additional likelihoods. We first added SPT/ACT (HighL), which didn't change the values of the parameters much and didn't reduce the error bars either. Adding in addition BAO reduced the error bars and shifted the values of the parameters back to the values prior to the introduction of cosmic strings and $N_{eff}$. This however reduced the allowed contribution from cosmic strings as well.

In terms of the degeneracies that appear, the most interesting case is the one with $N_{eff}$ and tensor modes. In this case the data suggest as a best fit $r=0.12 \pm 0.09$, so it is non-zero at $1\sigma$ level. The Hubble constant is increased as well to $H_0=80.59 \pm 6.57$. This is in fact due to the degeneracies introduced by cosmic strings, which are illustrated in Figure \ref{neff_r_planck2}. The allowed value of the string tension is quite large as well, $G\mu/c^2<2.49 \times 10^{-7}$. This degeneracies disappear however after adding BAO, reducing the cosmic string contribution to $G\mu/c^2<1.69 \times 10^{-7}$ at 95\% confidence level.

The same situation is true when adding other parameters such as running parameter in addition to $N_{eff}$ and strings. This also allows the contribution from cosmic strings to increase, up to $G\mu/c^2=2.49 \times 10^{-7}$ in the $N_{eff}$ \& r option. The results with just $N_{eff}$ added (and no strings) can be restored by adding HighL and BAO data. In this case, from the one-dimensional likelihood plots we see that the cosmic strings contribution is reduced, but a non-zero value is favoured (Figure \ref{planck_1dlike}). The values of the cosmological parameters which became very large drop considerably after adding the SPT/ACT likelihoods and the BAO. This one-dimensional plot is illustrative for the influence of the BAO in returning cosmological parameters close to their standard $\Lambda$CDM + $N_{eff}$ values by suppressing degeneracies. Hence, in the simplest $\Lambda$CDM \& strings model, the degeneracies between cosmic strings and other parameters are small, and BAO has a very small influence on cosmic strings. The same outcome appears when we additionally add tensor modes. For all the scenarios with $N_{eff}$ the values of the parameters increase massively after adding cosmic strings. The case with $N_{eff}$, $r$ and strings is a bit different compared to the others allowing additional degrees of freedom, in the sense that after adding BAO the contribution from cosmic strings is again consistent with zero, just as in the $\Lambda$CDM + strings scenario. The process of the increase of the values of the parameters in a scenario with additional degrees of freedom is illustrated in Figure \ref{planck_2d}.

From the  cases listed in the Appendix (Table \ref{params_planck_appendix}), the most interesting case from the cosmic strings point of view is the one with strings, $r$, $n_{run}$ and $m_{\nu,{\rm{sterile}}}^{\rm{eff}}$. In this scenario, $G\mu/c^2<2.57 \times 10^{-7}$ at 95\% confidence level and also $\Delta N_{eff}>0$ at $2\sigma$ level. The Hubble constant is only slightly larger than the $N_{eff}$ only value.

The degeneracies between $N_{eff}$ and cosmic strings have been studied in the context of the Abelian-Higgs cosmic string model \cite{PhysRevD.86.123014} and the authors have obtained a similar conclusion.

\section{BICEP2 case}

\begin{figure}[!h]
\begin{center}$
\begin{array}{cc}
\includegraphics[width=2.3in]{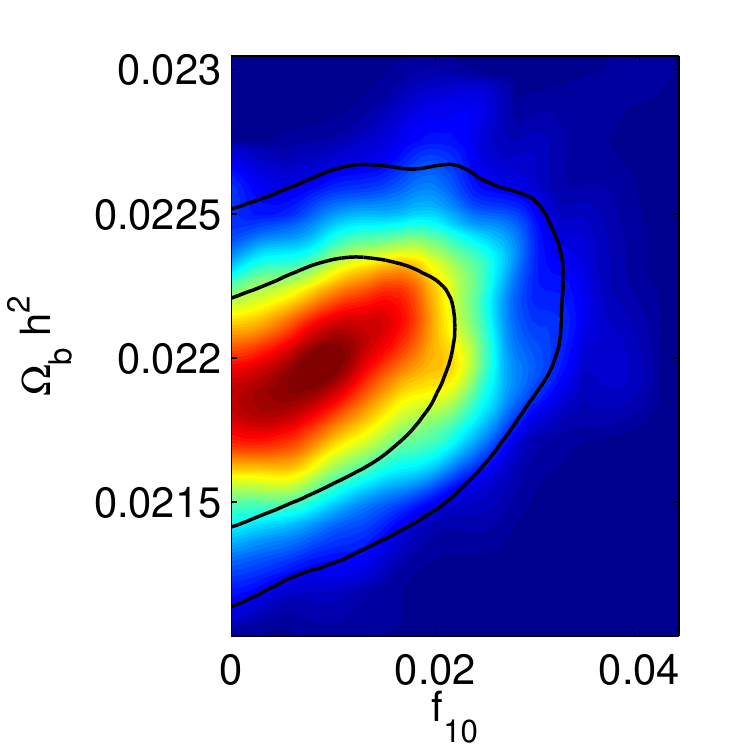} \\
\includegraphics[width=2.8in]{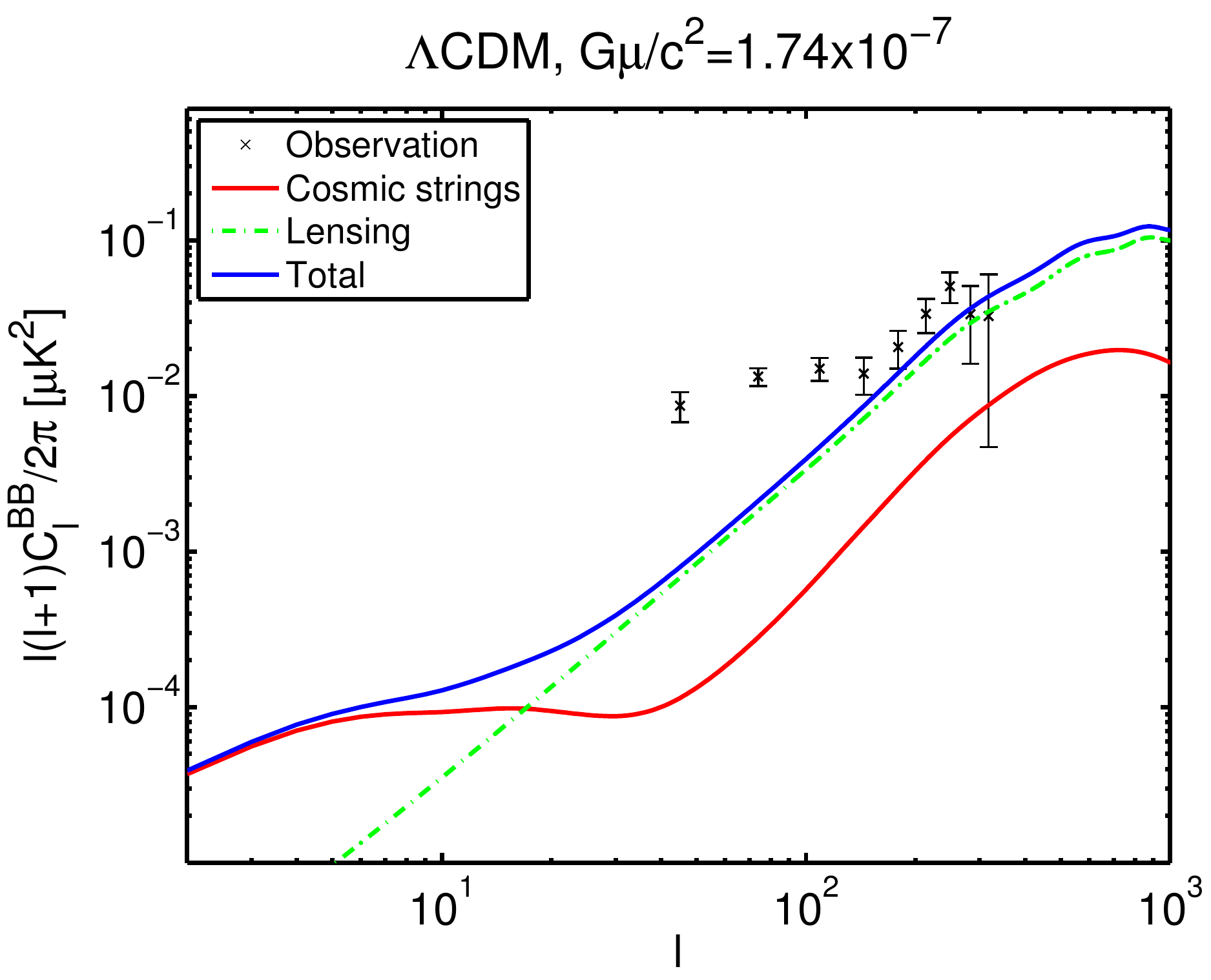} 
\end{array}$
\caption{Marginalized likelihoods in the $f_{10}$-$\Omega_bc^2$ plane (left) and best fit of the BB power spectrum using the Planck and BICEP2 likelihoods with cosmic strings at $G\mu=1.74 \times 10^{-7}$ (right) and $\Lambda$CDM}
\label{BB_direct}
\end{center}
\end{figure}

We have tried to explain the recently released BICEP2 data using Nambu-Goto cosmic strings. However, due to the amplitude of the signal, if one would try to fit the data only with cosmic strings would require a tension of $G\mu = 8.8 \times 10^{-7}$. Such a high value of $G\mu$ is not allowed by the stronger constraints from the TT power spectrum \cite{planckstr} and is at the limit of the constraints from the bispectrum. However, with the new BICEP2 data the allowed contribution from the cosmic strings is increased compared to using Planck data alone, because these are a source of BB polarisation. Using full likelihood calculations, we have found an increase of about 16\% in the string tension by adding the BICEP2 likelihoods compared to using only Planck data (but without adding additional parameters), to $G\mu$ to $1.74 \times 10^{-7}$. The other cosmological parameters are not significantly affected by the inclusion of cosmic strings (see later), but the BB power spectrum is not fitted very well, as it can be observed in Figure \ref{BB_direct}.

We have considered various possibilities of fitting the data without tensor modes ($r=0$), but the fit values did not improve. The easiest option was to introduce the tensor modes together with cosmic strings. In that situation we have obtained a value of $r=0.15 \pm 0.04$ and $G\mu<1.44 \times 10^{-7}$ with $f_{10}<0.026$ at 95\% confidence level. In this case, the tensor-to-scalar ratio is decreased compared to the best-fit obtained by the BICEP2 team, but it is closer to value they have obtained after subtracting the dust foregrounds. There is no sign of degeneracy with any of the parameters.

\begin{figure}[h]
\begin{center}$
\begin{array}{c}
\includegraphics[width=3.5in]{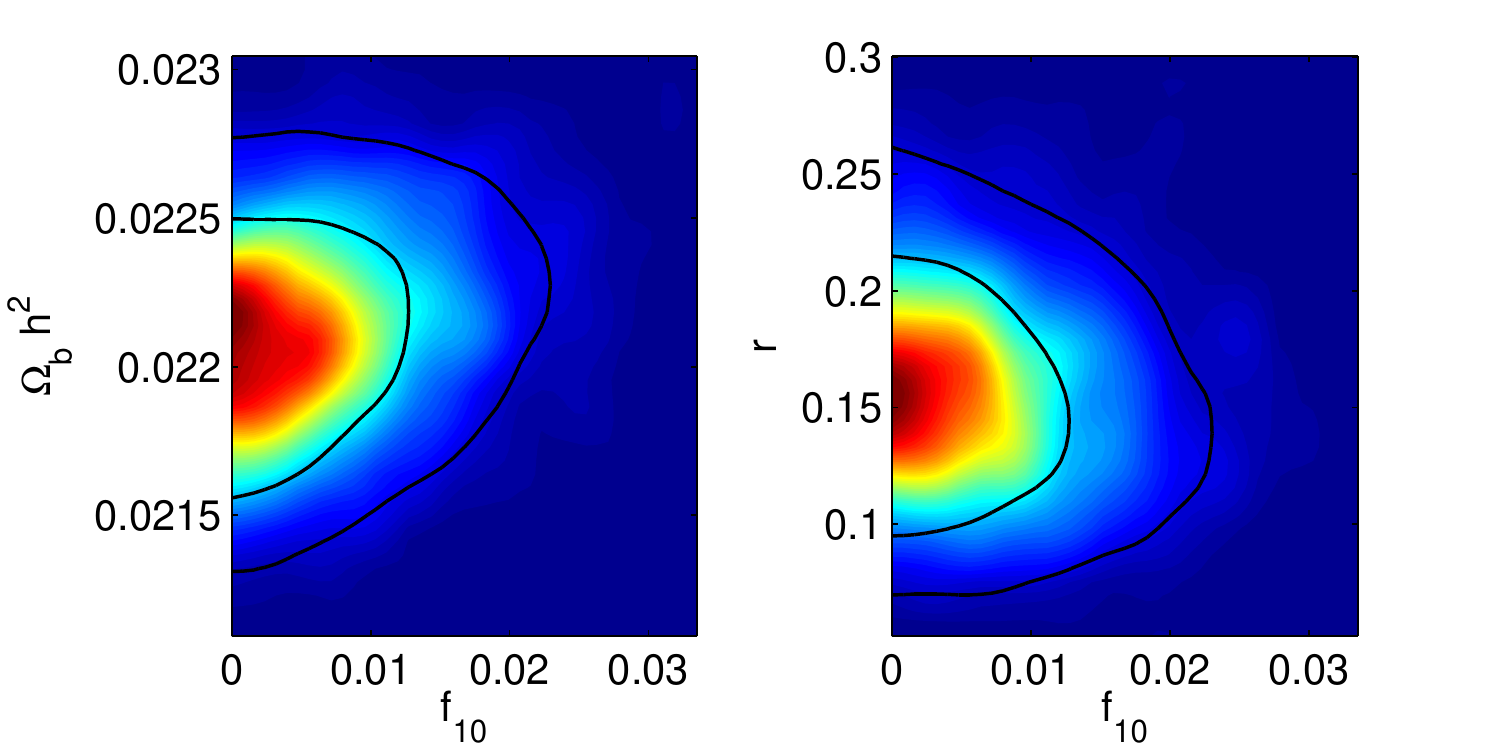} \\ 
\includegraphics[width=2.8in]{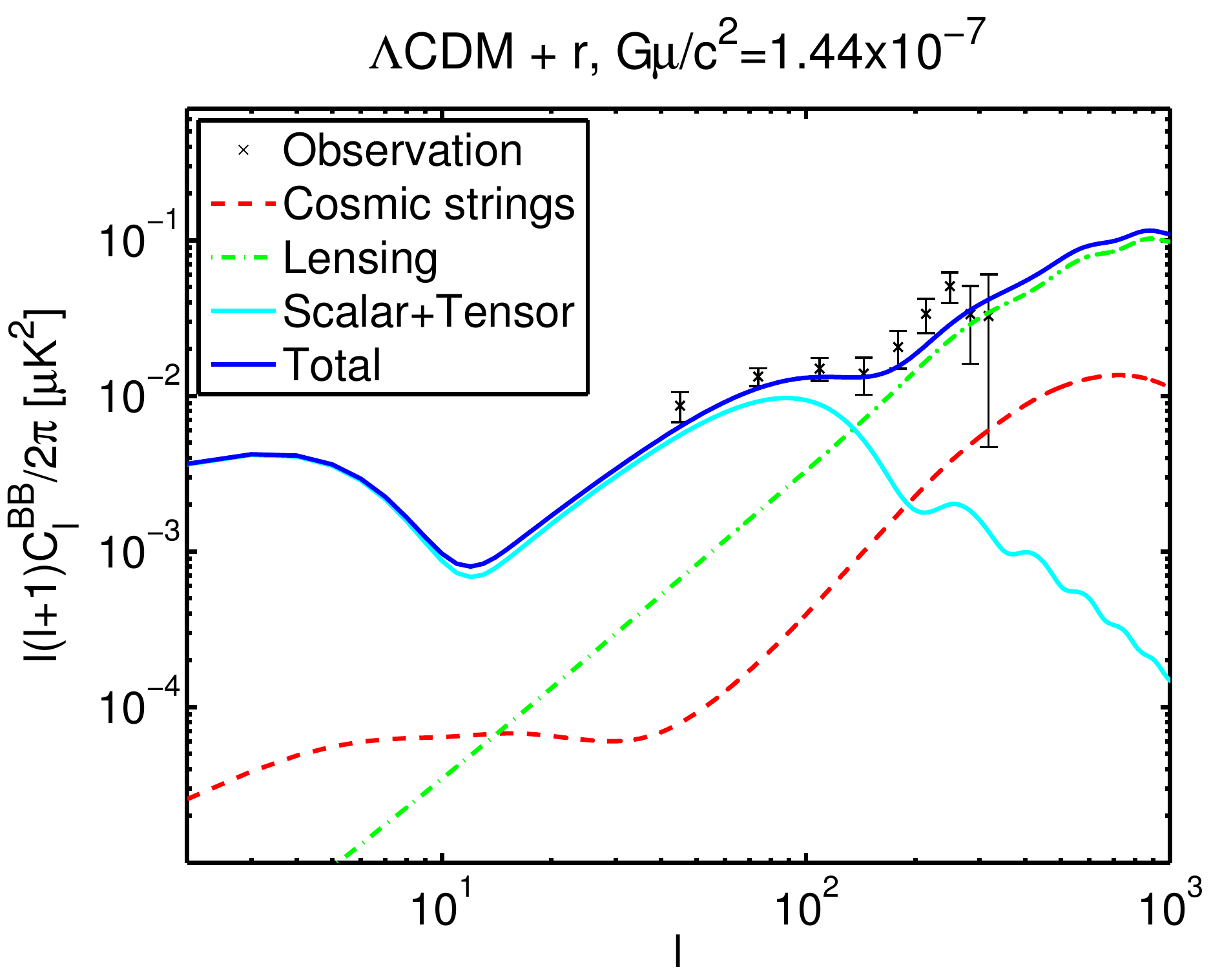}
\end{array}$
\caption{Marginalised likelihoods in the $f_{10}$-$\Omega_bc^2$ and $f_{10}$-$r$ planes for BICEP2 likelihoods with strings and tensor modes (top) and best fit of the BB power spectrum using the Planck and BICEP2 likelihoods with cosmic strings at $G\mu=1.44 \times 10^{-7}$ (right) and $\Lambda$CDM and $r$}
\label{r_only}
\end{center}
\end{figure}

\begin{figure}[!htb]
\begin{center}$
\begin{array}{c}
\includegraphics[width=3.6in]{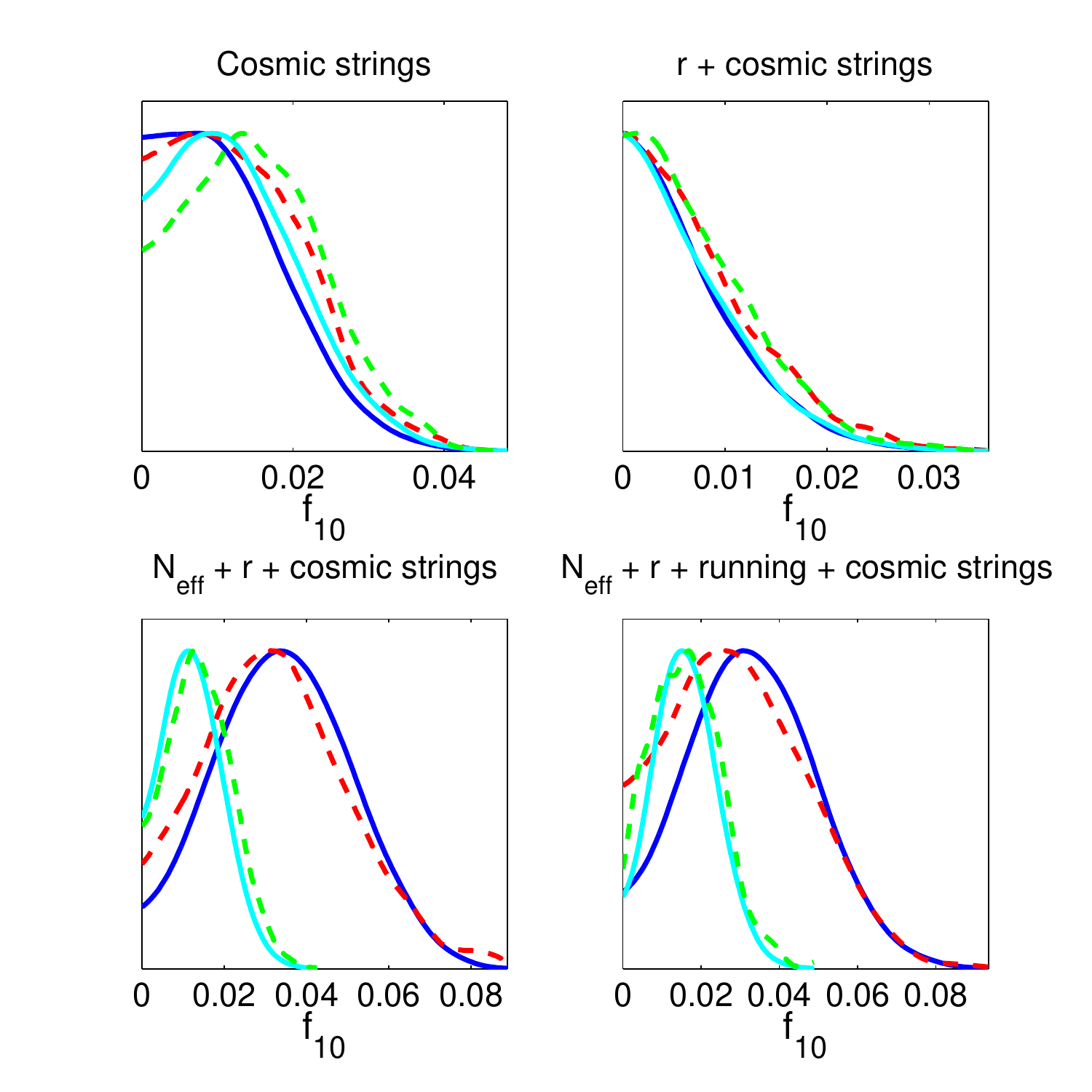} 
\end{array}$
\caption{Mean likelihoods of the samples (red dotted lines) and marginalised probabilities (blue solid lines) for parameter $f_{10}$ in the following situations (from left to right and top to bottom): Planck \& BICEP \& strings; Planck \& BICEP \& strings \& r; Planck \& BICEP \& strings \& $N_{eff}$ \& $r$, Planck \& BICEP \& $N_{eff}$ \& running \& r \& cosmic strings. The green and cyan curves respectively represent the mean likelihoods and marginalised probabilities of the samples after the introduction of HighL \& BAO.}
\label{bicep_1dlike}
\end{center}
\end{figure}

We note that in the absence of tensor modes a non-zero contribution from cosmic strings is favoured, but this disappears as soon as $r$ is introduced. This is due to the fact that although cosmic strings cannot explain the BB polarisation signal by having the wrong shape (even if we allow arbitrary large $G\mu/c^2$) they are still able to help fitting the BICEP2 data point in the absence of tensor modes. As the tensor modes are introduced, they take over the string contribution by giving the correct shape in the polarisation domain. This is illustrated clearly by the one-dimensional likelihood plots for $f_{10}$ on the first row of Figure \ref{bicep_1dlike}. Baryon acoustic oscillations do not change the result significantly in the case with tensors. However, for the string-only one, they fix the contribution from strings to a non-zero value.

The most interesting cases are described in Table \ref{params_bicep}, while the other cases that we have considered are listed in the Appendix (Table \ref{params_bicep_appendix}). As in the Planck case, interesting  degeneracies appear due to $N_{eff}$ and a similar outcome can also be observed. 

\begin{table*}[p!]

\centering
\caption{Values of the cosmological parameters in the Planck + BICEP2 likelihoods case, with $1\sigma$ error levels full likelihood analysis (all cases also include the Planck nuisance parameters)}
\begin{tabular}{|c|C{3cm}|C{3cm}|C{3cm}|C{3cm}|}
\hline
Parameter                           & $\Lambda$CDM        & strings            & $r$                 & strings, $r$            \\ \hline
$G\mu/c^2 < (2\sigma)$              & -                   & $1.74 \times 10^{-7}$& - & $1.44 \times 10^{-7}$ \\ \hline
$G\mu/c^2$ (best fit)               & -                   & $8.46 \times 10^{-8}$& - & $8.30 \times 10^{-8}$ \\ \hline
$n_{run}$                           & -                   & -                  &  -& -                   \\ \hline
$r$                                 & -                   & -                  & $0.16 \pm 0.04$    & $0.15 \pm 0.04$                   \\ \hline
$\Delta N_{eff}$                    & -                   & -                  & -                  & -       \\ \hline 
$H_0$                               & $66.26 \pm 1.15$    & $66.76 \pm 1.20$   & $67.72 \pm 1.10$   & $67.95 \pm 1.20$       \\ \hline 
$100\Omega_b h^2$                   & $2.183 \pm 0.27$    & $2.197 \pm 0.030$  & $2.203 \pm 0.028$  & $2.210 \pm 0.029$    \\ \hline
$\Omega_c h^2$                      & $0.122 \pm 0.003$   & $0.121 \pm 0.003$  & $0.119 \pm 0.003$  & $0.118 \pm 0.003$    \\ \hline
$\tau$                              & $0.093 \pm 0.013$   & $0.090 \pm 0.013$  & $0.089 \pm 0.013$  & $0.088 \pm 0.013$   \\ \hline
$100\theta_{MC}$                    & $1.041 \pm 0.0006$  & $1.0411 \pm 0.0007$ & $1.0413 \pm 0.0006$ & $1.0414 \pm 0.0007$    \\ \hline
$ln(10^{10}A_s)$                    & $3.101 \pm 0.00255$ & $3.084 \pm 0.027$  & $3.085 \pm 0.025$  & $3.075 \pm 0.025$   \\ \hline
$n_s$                               & $0.954 \pm 0.0070$  & $0.953 \pm 0.007$  & $0.964 \pm 0.007$   & $0.964 \pm 0.007$     \\ \hline
$-ln\mathcal{L}$                    & 4946.7              & 4946.1             & 4926.5              & 4926.6              \\ \hline \hline

Parameter                           & $N_{eff}$, $r$        & $N_{eff}$, $r$, strings (Planck) & $N_{eff}$, $r$, strings (Planck + HighL) & $N_{eff}$, $r$, strings (Planck + HighL + BAO)            \\ \hline
$G\mu/c^2 < (2\sigma)$              & -                   & $2.72 \times 10^{-7}$ & $2.47 \times 10^{-7}$ & $1.70 \times 10^{-7}$ \\ \hline
$G\mu/c^2$ (best fit)               & -                   & $1.96 \times 10^{-7}$& $1.68 \times 10^{-7}$ & $1.43 \times 10^{-7}$ \\ \hline
$n_{run}$                           & -                   & -                  & -                  & -                   \\ \hline
$r$                                 & $0.20 \pm 0.05$     &  $0.23 \pm 0.06$   & $0.22 \pm 0.06$    & $0.16 \pm  0.04$      \\ \hline
$\Delta N_{eff}$                    & $1.1061 \pm 0.42$   & $2.19 \pm 0.69$    & $2.234 \pm 0.637$    & $0.908 \pm 0.331$      \\ \hline 
$H_0$                               & $76.14 \pm 3.52$    & $85.86 \pm 6.06$   & $85.72 \pm 5.42$  & $73.21 \pm 2.11$    \\ \hline 
$100\Omega_b h^2$                   & $2.287 \pm 0.044$   & $2.408  \pm 0.073$  & $ 2.398 \pm 0.062$ & $2.255 \pm  0.031$ \\ \hline
$\Omega_c h^2$                      & $0.132 \pm 0.006$   & $0.141 \pm 0.007$  & $ 0.143 \pm  0.007$& $0.133 \pm 0.006$ \\ \hline
$\tau$                              & $0.101 \pm 0.015$   & $0.135 \pm 0.016$  & $ 0.104 \pm 0.016$ & $0.089 \pm 0.013$  \\ \hline
$100\theta_{MC}$                    & $1.0403 \pm 0.0007$ & $1.0400 \pm 0.0007$ & $1.0399 \pm 0.0007$ & $1.0401 \pm  0.0007$  \\ \hline
$ln(10^{10}A_s)$                    & $3.136 \pm 0.033$   & $3.132 \pm 0.034$  & $3.137 \pm 0.034$  & $3.107 \pm  0.028$    \\ \hline
$n_s$                               & $1.006 \pm 0.017$   & $1.039 \pm 0.023$ & $1.037 \pm 0.022$  & $0.986 \pm  0.011$   \\ \hline
$-ln\mathcal{L}$                    & 4923.15             & 4922.6             & 5275.4             &  5279.0             \\ \hline  \hline

Parameter                           & $n_{run}$, $r$         & $r$, $n_{run}$, strings& $r$, $n_{run}$, $N_{eff}$, strings& $r$, $n_{run}$, $N_{eff}$, strings (Planck + HighL + BAO)               \\ \hline
$G\mu/c^2 < (2\sigma)$              & -                   & $2.07 \times 10^{-7}$& $2.65 \times 10^{-7}$& $1.88 \times 10^{-7}$\\ \hline
$G\mu/c^2$ (best fit)               & -                   & $9.42 \times 10^{-8}$& $1.28 \times 10^{-7}$& $1.08 \times 10^{-7}$ \\ \hline
$n_{run}$                           & $-0.028 \pm 0.010$  & $-0.036 \pm 0.011 $ & $-0.012 \pm 0.02$& $-0.031 \pm 0.012$ \\ \hline
$r$                                 & $0.19 \pm 0.04$     & $0.22 \pm 0.05$    & $0.23 \pm 0.06$ & $0.22 \pm 0.05$ \\ \hline
$\Delta N_{eff}$                    & -                   & -& $1.426 \pm 0.927$& $0.448 \pm 0.310$\\ \hline 
$H_0$                               & $67.72 \pm 1.23$    & $68.27 \pm 1.28$    & $82.03 \pm 7.61$& $70.79 \pm 1.98$ \\ \hline
$100\Omega_b h^2$                   & $2.234 \pm 0.315$   & $2.262 \pm 0.037$   & $2.384 \pm 0.077$& $2.271 \pm 0.031$ \\ \hline
$\Omega_c h^2$                      & $0.119 \pm 0.003$   & $0.119 \pm 0.003$   & $0.136 \pm 0.010$& $0.126\pm 0.005$ \\ \hline
$\tau$                              & $0.104 \pm 0.016$   & $0.103 \pm 0.016$   & $0.105 \pm 0.017$& $0.102 \pm 0.1053$ \\ \hline
$100\theta_{MC}$                    & $1.0414 \pm 0.0007$ & $1.0420 \pm 0.0007$ & $1.0403 \pm 0.0008 $ & 1.0409$ \pm 0.0007$ \\ \hline
$ln(10^{10}A_s)$                    & $3.121 \pm 0.031$   & $3.108 \pm 0.033$   & $3.131 \pm 0.035$& $3.121 \pm 0.032$ \\ \hline
$n_s$                               & $0.958 \pm 0.008$   & $0.954 \pm 0.008$   & $1.020 \pm 0.033$& $0.967 \pm 0.012$ \\ \hline
$-ln\mathcal{L}$                    & 4922.9              & 4922.8              & 4922.0           & 5276.5  \\ \hline
\end{tabular}
\label{params_bicep}
\end{table*}

\begin{figure*}[!p]	
\begin{center}$
\begin{array}{cc}
\includegraphics[width=0.77in]{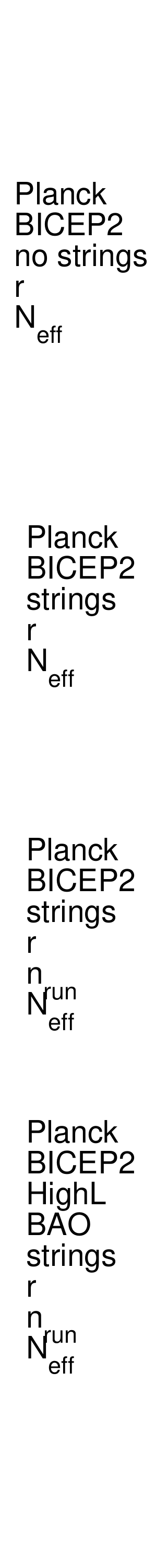}
\includegraphics[width=6.05in]{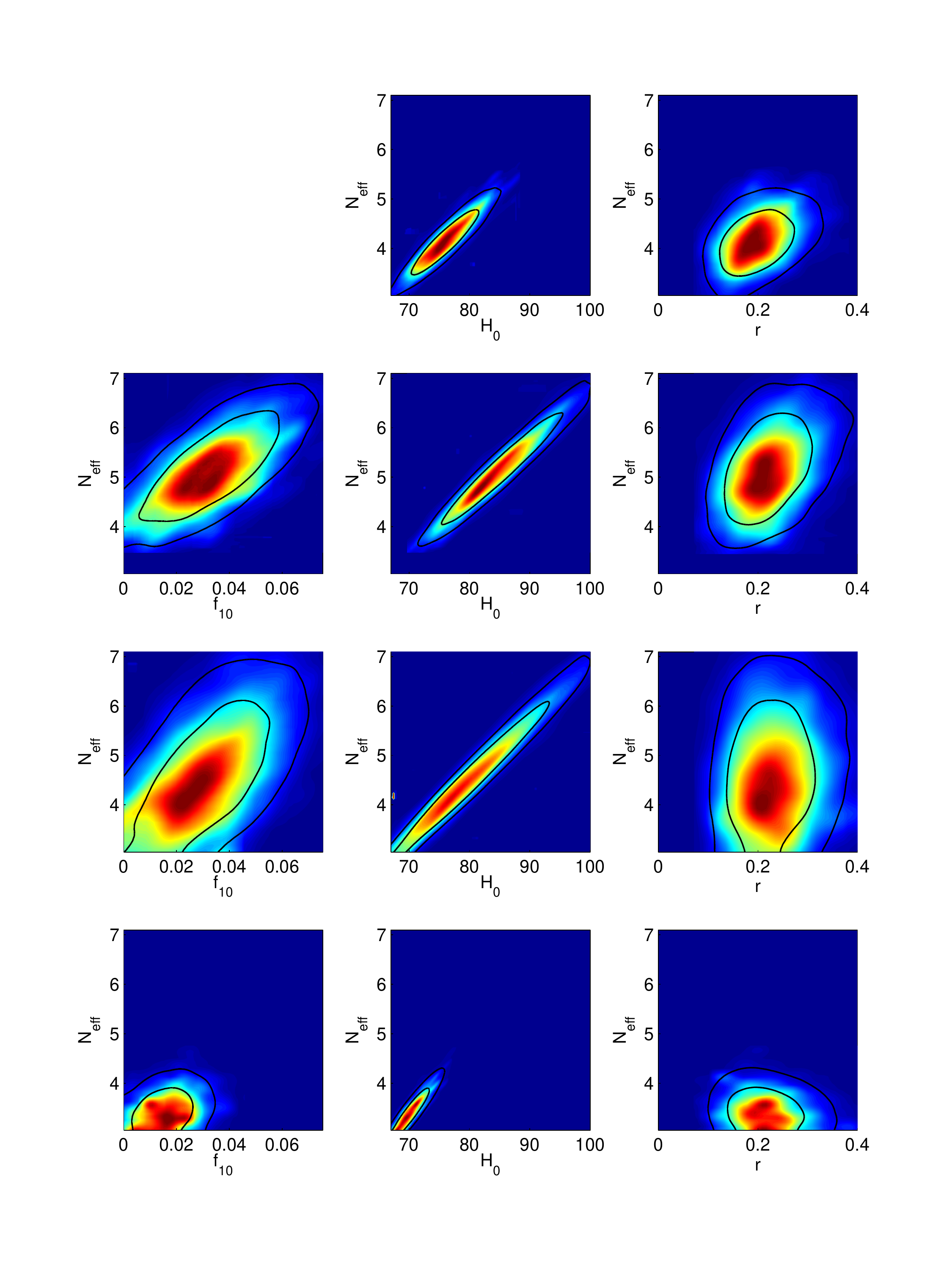}
\end{array}$
\caption{Two-dimensional marginalised likelihoods in the $f_{10}$-$N_{eff}$, $H_0$-$N_{eff}$ and $n_s$-$N_{eff}$ planes in the following cases (top to bottom): $N_{eff}$ only (no strings), $N_{eff}$ and cosmic strings, $N_{eff}$, running and cosmic strings, $N_{eff}$, running and cosmic strings, with SPT/ACT and BAOs}
\label{bicep_2d}
\end{center}
\end{figure*}

The cosmic strings contribution is very large with the  Planck \& BICEP2 likelihoods, but the Hubble constant and $N_{eff}$ are also very big (see Table \ref{params_bicep} and Figure \ref{bicep_2d}). Adding ACT/SPT and BAO recovers the $\Lambda$CDM values for the cosmological parameters and reduces the contribution from cosmic strings. Nevertheless, a non-zero contribution is still preferred (bottom row of Figure \ref{bicep_1dlike}). In both the scenarios of cosmic strings \& $N_{eff}$ \& $r$ and cosmic strings \& $N_{eff}$ \& $r$ \& $n_{run}$ the preferred value of $f_{10}$ is non-zero and the distribution is wide. Adding SPT/ACT and BAO reduces the preferred value for $f_{10}$, but also narrows the distribution. This can be compared to Figure \ref{planck_1dlike}, but here the BICEP2 polarisation data favours more a non-zero contribution of cosmic strings.

From the scenarios listed in Table \ref{params_bicep_appendix} in the Appendix, the one which allows the most cosmic strings is the one with just $N_{eff}$ and $m_{\nu,{\rm{sterile}}}^{\rm{eff}}$. Indeed, in this case the best fit value is $G\mu/c^2=2.36 \times 10^{-7}$ and the constraint is  $G\mu/c^2<2.99 \times 10^{-7}$ at $2\sigma$ level. Both the Hubble constant and $N_{eff}$ have big values and the fit is not very good not having tensor modes.

\section{Conclusions}

We have used the power spectrum obtained from high-resolution Nambu-Goto cosmic string simulations \cite{string-paper1} to constrain the string tension magnitude by adding cosmic strings to the standard 6-parameter $\Lambda$CDM model \cite{planckres}. In this simplest model, we obtained a string tension constraint of $G\mu/c^2<1.49 \times 10^{-7}$ (95\% confidence), using the Planck likelihood and WMAP polarisation. This result is comparable to the value obtained by the Planck team \cite{planckstr}. In this case, the string tension $G\mu/c^2$ does not introduce extra degeneracies between $\Lambda$CDM parameters. However, by allowing $N_{eff}$ to vary, the string constraint gets much weaker ($G\mu/c^2<2.28 \times 10^{-7}$) and the Hubble constant increases to $H_0=75.96$, with a significant degeneracy between $f_{10}$ and $H_0$. This degeneracy disappears however by adding BAOs and HighL contributions. In that case, the string tension reverts close to its previous value,  $G\mu/c^2<1.58 \times 10^{-7}$. We note that BAOs are the key ingredient for breaking the degeneracies as HighL data cannot alone solve the problem. The same behaviour is observed when allowing tensor modes in addition to $N_{eff}$, where  the constraint on $G\mu/c^2$ shifts from $2.49 \times 10^{-7}$ to $1.56 \times 10^{-7}$ with BAOs. These degeneracies can be more easily interpreted visually (see Figs. \ref{planck_direct} and \ref{neff_r_planck2}). By adding running in addition to cosmic strings, the string constraint becomes slightly weaker ($G\mu/c^2<1.88 \times 10^{-7}$), but does not induce significant degeneracies. In addition, we have also analysed the contribution of an additional sterile neutrino and we have found no significant differences to the parameter values.

We have performed a similar analysis by considering the BICEP2 data in addition to Planck likelihoods and WMAP polarisation. In this case, the string tension constraints loosen, but the new polarisation signal cannot be explained solely by cosmic strings with no contribution from primordial tensor modes. This is due to the fact that cosmic strings are tightly constrained by the temperature data. Hence, in a pure $\Lambda$CDM and strings scenario, the 95\% confidence level constraint on the string tension only rises to $G\mu/c^2<1.74 \times 10^{-7}$ (Figure \ref{BB_direct}). By adding tensor modes, we note that the model prefers a value of $r=0.15$ and $G\mu/c^2<1.44 \times 10^{-7}$ and strings are not favoured. Adding additionally $N_{eff}$ greatly increases the allowed amount of cosmic strings to $2.72 \times 10^{-7}$, but the values of $r$, of the Hubble constant and of $\Delta N_{eff}$ are increased as well, 0.20 to 85.86 and 2.19 respectively. This is due to the same degeneracies that appear. BAOs and SPT/ACT likelihoods again revert the situation to $\Lambda$CDM with $G\mu/c^2<1.70 \times 10^{-7}$, $r=0.16$ $\Delta N_{eff}=0.908$ and $H_0=73.21$. We note again that the SPT/ACT likelihoods make little difference to the results (Table \ref{params_bicep}). By also allowing for a non-zero running of the spectral index we see from Table \ref{params_bicep} that running in itself allows for more cosmic strings ($G\mu/c^2<2.07 \times 10^{-7}$) and tensor modes ($r=0.22$) but the degeneracies are modest and the Hubble constant keeps its usual value ($H_0=68.27$). $N_{eff}$, when added to this model, induces huge degeneracies and shifts the Hubble parameter but again this problem is cured with BAOs. 

We anticipate using our improved cosmic string predictions with the second Planck Data Release shortly.

\begin{acknowledgments}
The authors are very grateful to Martin Landriau, Carlos Martins and Levon Pogosian for many enlightening discussions. AL is supported by STFC. This work was supported by an STFC consolidated grant ST/L000636/1. This work was undertaken on the COSMOS Shared Memory system at DAMTP, University of Cambridge operated on behalf of the STFC DiRAC HPC Facility. This equipment is funded by BIS National E-infrastructure capital grant ST/J005673/1 and STFC grants ST/H008586/1, ST/K00333X/1.
\end{acknowledgments}

\section{Appendix}
In this Appendix we list the Tables with results that were not included in the main part of the paper, in the Planck \& WP and Planck \& WP \& BICEP2 scenarios.

\begin{table*}[p!]
\centering
\caption{Values of the cosmological parameters when considering only the Planck and WP likelihoods, with $1\sigma$ error levels full likelihood analysis}
\begin{tabular}{|c|C{3cm}|C{3cm}|C{3cm}|C{3cm}|}
\hline

Parameter                           & strings, $r$          & strings, $r$, running& $N_{eff}$ (Planck + HighL) & $N_{eff}$ (Planck + HighL + BAO) \\ \hline
$G\mu/c^2 < (2\sigma)$              & $1.42 \times 10^{-7}$ & $1.99 \times 10^{-7}$  & - & - \\ \hline  
$G\mu/c^2$ (best fit)               & $5.09 \times 10^{-8}$ & $8.39 \times 10^{-9}$  & - & - \\ \hline   
$n_{run}$                           & -                   & $-0.029 \pm 0.012$ & -                   & -                   \\ \hline
$r$                                 & $0.039 \pm 0.036$   & $0.11 \pm 0.09$    & -                   & -                \\ \hline    
$\Delta N_{eff}$                    & -                   & -                  & $0.669 \pm 0.323 $ & $0.531 \pm 0.255$ \\ \hline
$H_0$                               & $67.59 \pm 1.21$    & $67.92 \pm 1.31$   & $71.86 \pm 2.76$   & $70.61 \pm 1.68 $ \\ \hline
$100\Omega_b h^2$                   & $2.211 \pm 0.029$   & $2.254 \pm 0.038$  & $2.241 \pm 0.038$  & $2.227 \pm 0.028$ \\ \hline
$\Omega_c h^2$                      & $0.119 \pm 0.003$   & $0.119 \pm 0.003$  & $0.129 \pm 0.005$  & $0.128 \pm 0.004$ \\ \hline
$\tau$                              & $0.087 \pm 0.013$   & $0.101 \pm 0.016$  & $0.095 \pm 0.015$  & $0.092 \pm 0.013$ \\ \hline
$100\theta_{MC}$                    & $1.0413 \pm 0.0006$ & $1.0415\pm 0.0007$ & $1.0404 \pm 0.0007$& $1.0405\pm 0.0007$ \\ \hline
$ln(10^{10}A_s)$                    & $3.077 \pm 0.025$   & $3.105 \pm 0.032$  & $3.120 \pm 0.032$  & $3.112 \pm 0.027$ \\ \hline
$n_s$                               & $0.960 \pm 0.007$   & $0.953 \pm 0.009$  & $0.980 \pm 0.014$  & $0.974 \pm 0.010$ \\ \hline
$-ln\mathcal{L}$                    & 4903.2              &  4902.0            & 5255.3             &  5258.58           \\ \hline \hline         

Parameter                           & $N_{eff}$,  $m_{\nu,{\rm{sterile}}}^{\rm{eff}}$ & strings, $N_{eff}$, $m_{\nu,{\rm{sterile}}}^{\rm{eff}}$ & strings, $r$, $N_{eff}$,  $m_{\nu,{\rm{sterile}}}^{\rm{eff}}$ & strings, $N_{eff}$, $r$ (Planck + HighL + BAO) \\ \hline 
$G\mu/c^2 < (2\sigma)$              & -                   & $2.36 \times 10^{-7}$ & $2.57 \times 10^{-7}$ & $1.56 \times 10^{-7}$ \\ \hline
$G\mu/c^2$ (best fit)               & -                   & $9.17 \times 10^{-8}$ & $1.44 \times 10^{-7}$ &\\ \hline
$r$                                 & -                   & -                    & $0.11 \pm 0.90$   & $0.05 \pm 0.04$   \\ \hline
$\Delta N_{eff}$                    & $0.535 \pm 0.306$   & $1.055 \pm 0.535$    & $1.522 \pm 0.725$ & $0.696 \pm 0.308$    \\ \hline  
$m_{\nu,{\rm{sterile}}}^{\rm{eff}}$ [eV] & $0.261 \pm 0.222$& $0.36 \pm 0.32$    & $0.38 \pm 0.35$   & -                \\ \hline 
$H_0$                               & $69.14 \pm 2.74$    & $72.71 \pm 4.79$     & $76.69 \pm 6.48$  & $71.79 \pm 2.00$\\ \hline   
$100\Omega_b h^2$                   & $2.233 \pm 0.035$   & $2.293 \pm 0.058$    & $2.338  \pm 0.078$& $2.244 \pm 0.030$ \\ \hline
$\Omega_c h^2$                      & $0.127 \pm 0.005$   & $0.132 \pm 0.007$    & $0.136 \pm 0.008$ & $0.130 \pm 0.005$ \\ \hline
$\tau$                              & $0.095\pm 0.015$    & $0.097 \pm 0.015$    & $0.099 \pm 0.016$ & $0.089 \pm 0.012$\\ \hline
$100\theta_{MC}$                    & $1.0405\pm 0.0007$  & $1.0402\pm 0.0008$   & $1.0399\pm 0.0008$& $1.0403 \pm 0.0007$  \\ \hline
$ln(10^{10}A_s)$                    & $3.117\pm 0.032$    & $3.111 \pm 0.033$    & $3.120\pm 0.027$  & $3.103 \pm 0.028$ \\ \hline
$n_s$                               & $0.975 \pm 0.014$   & $0.989 \pm 0.020$    & $1.010\pm 0.027$  & $0.978 \pm 0.010$ \\ \hline
$-ln\mathcal{L}$                    & 4902.5              & 4902.6               & 4902.3   &  5259.6   \\ \hline  
\end{tabular}
\label{params_planck_appendix}
\end{table*}

\begin{table*}[p!]
\centering
\caption{Values of the cosmological parameters in the Planck + WP + BICEP2 likelihoods case, with $1\sigma$ error levels full likelihood analysis}
\begin{tabular}{|c|C{3cm}|C{3cm}|C{3cm}|C{3cm}|}
\hline
Parameter                           & $n_{run}$, strings  & $r$, $N_{eff}$, $n_{run}$, strings (Planck + HighL)  & $N_{eff}$, strings, $m_{\nu,{\rm{sterile}}}^{\rm{eff}}$ & $r$, $N_{eff}$, strings, $m_{\nu,{\rm{sterile}}}^{\rm{eff}}$               \\ \hline
$G\mu/c^2 < (2\sigma)$              & $2.25 \times 10^{-7}$ & $2.41 \times 10^{-7}$& $2.99 \times 10^{-7}$ & $2.85 \times 10^{-7}$\\ \hline
$G\mu/c^2$ (best fit)               & $1.56 \times 10^{-8}$ & $1.96 \times 10^{-7}$& $2.36 \times 10^{-7}$ & $1.53 \times 10^{-7}$\\ \hline
$n_{run}$                           & $-0.025 \pm 0.010$  & $-0.012 \pm 0.017$ & -& - \\ \hline
$r$                                 & -                   & $0.23 \pm 0.06$    & -& $0.22 \pm 0.06$ \\ \hline
$\Delta N_{eff}$                    & -                   & $1.703 \pm 0.910$ & $1.87 \pm 0.67$ & $2.21 \pm 0.073$\\ \hline 
$m_{\nu,{\rm{sterile}}}^{\rm{eff}}$ [eV] & -              & -            &$0.20 \pm 0.19$ & $0.30 \pm 0.25$\\ \hline
$H_0$                               & $66.92 \pm 1.28$    & $81.61 \pm 7.31$    & $80.00 \pm 5.92$& $83.15 \pm 6.62$  \\ \hline
$100\Omega_b h^2$                   & $2.236 \pm 0.036$   & $2.373 \pm 0.067$   & $2.369 \pm 0.072$& $2.402 \pm 0.076$ \\ \hline
$\Omega_c h^2$                      & $0.122 \pm 0.003$   & $0.137 \pm 0.010$   & $0.142 \pm 0.008$& $0.142 \pm 0.008$ \\ \hline
$\tau$                              & $0.100 \pm 0.016$   & $0.106 \pm 0.016$   & $0.103 \pm 0.016$& $0.103 \pm 0.016$ \\ \hline
$100\theta_{MC}$                    & $1.0413 \pm 0.0007$ & $1.0403 \pm 0.0008$ & $1.0397 \pm 0.0007$& $1.0397 \pm 0.0007$ \\ \hline
$ln(10^{10}A_s)$                    & $3.104 \pm 0.033$   & $3.137 \pm 0.034$   & $3.127 \pm 0.035$& $3.129 \pm 0.034$ \\ \hline
$n_s$                               & $0.944 \pm 0.008$   & $1.017 \pm 0.033$   & $1.013 \pm 0.022$& $1.035 \pm 0.024$ \\ \hline
$-ln\mathcal{L}$                    & 4943.6              & 5275.7              & 4943.1& 4922.2 \\ \hline
\end{tabular}
\label{params_bicep_appendix}
\end{table*}

\clearpage
\bibliography{Bibliografie}{}

\end{document}